\documentclass[aps,preprint,showpacs, showkeys]{revtex4}%
\usepackage{amsfonts}
\usepackage{amsmath}
\usepackage{amssymb}
\usepackage{graphicx}%
\providecommand{\U}[1]{\protect\rule{.1in}{.1in}}

\begin{document}
\title[ ]{Non-perturbative calculations for the effective potential of the $PT$ symmetric
and non-Hermitian $\left( -g\phi^{4}\right) $ field theoretic model}
\author{Abouzeid. M. Shalaby\thanks{E-mail:amshalab@ mans.eg.edu}}
\affiliation{Theoretical Physics Group, Physics Department, Faculty of Science, Mansoura University, Egypt
P.O. Box 35516, Egypt}

\begin{abstract}
We investigate the effective potential of the $PT$ symmetric $\left(  -g\phi^{4}\right)  $ field theory, perturbatively as well as non-perturbatively. For the perturbative calculations, we first use normal ordering to obtain the first order effective potential from which the predicted vacuum condensate vanishes exponentially as $G\rightarrow G^+$ in agreement with previous calculations. For the higher orders, we employed the invariance of the bare parameters under the change of the mass scale $t$ to fix the transformed form totally equivalent to the original theory. The form so obtained up to $G^3$ is new and shows that all the 1PI amplitudes are perurbative for both $G\ll 1$ and $G\gg 1$ regions. For the intermediate region, we modified the fractal self-similar resummation method to have a unique resummation formula for all $G$ values. This unique formula is necessary because the effective potential is the generating functional for all the 1PI amplitudes which can be obtained via $\partial^n E/\partial b^n$ and thus we can obtain an analytic calculation for the 1PI amplitudes. Again, the resummed from of the effective potential is new and  interpolates the effective potential between the perturbative regions. Moreover, the resummed effective potential agrees in spirit of previous calculation concerning bound states.     
\end{abstract}
\maketitle

\section{Introduction}

In recent years, it has been established that the \textit{PT }symmetric and
non-Hermitian quantum models have real and discrete spectra
\cite{bendg1,bend1,bend2,bend3,bend4,bend5,bend6,bend7,bend8,bend9,bend10,bend11,bend12,bend13,bend14}%
. This may draw the attention to the reinvestigation of the previously
rejected non-Hermitian models especially for the quantum field versions. For those theories, there
are some clues concerning the importance of the study of them. For instance, the very recent result concerning the existence of a non-Hermitian representation for Hermitian theory which has simpler
calculation \cite{bend2005}. Accordingly, one may aim to describe hadrons in a
simple way rather than the complicated \textbf{QCD} model by finding a non-Hermitian
representation for \textbf{QCD}. However, this is still a speculation and in this work
we concentrate on the study of the effective potential of a simple model but
exhibits non-trivial features. A class of simple but non-trivial quantum
mechanical models are given by%

\begin{equation}
H=p^{2}+x^{2}\left(  ix\right)  ^{\epsilon}\text{, \ \ \ }\epsilon>0\text{.}
\label{qwang}%
\end{equation}
All such models have real and positive spectra even in the case of
$\epsilon=2$. In fact, all the complex PT symmetric
Hamiltonians have real and positive spectra \cite{bend1}.

Unlike the quantum mechanical versions of the \textit{PT }symmetric models,
the quantum field versions of such theories still are not investigated in a
sizable way. Moreover, the quantum field Hamiltonians have interesting
properties. For instance, a simple model with Lagrangian density like
\begin{equation}
L=\frac{1}{2}\left(  \left(  \partial\phi\right)  ^{2}-m^{2}\phi^{2}\right)
+\frac{g}{4}\phi^{4}, \label{fielng}%
\end{equation}
exhibits asymptotic freedom as well as having bound states \cite{bound}. Also,
for certain range of the coupling values it has two-body bound state
(meson-like) and for another range it has three-body bound state
(baryon-like). Thus, it is concluded that this simple model has supersymmetric
features. This may give us a hope to describe strong interactions with Abelian
theories without the need for Glouns. Regardless of these legitimate hopes,
the PT symmetric and non-Hermitian theories, like any of the physically
acceptable models, deserve the employment of the usual machinery of
investigation under different conditions. For example, we need to know how
they behave at zero and non-zero temperatures, the presence of external
sources, extreme conditions, etc... . Since the effective potential serves as
the generating functional for all the one-particle irreducible (1PI) amplitudes,
it's investigation is the basic stone for all other discussions. Up to the best of our knowledge,
the effective potential of the non-Hermitian $\phi^{4}_{1+1}$ has never been obtained in a
form reliable in all regions of the coupling space. This situation is
partly due to being a new field of study  and
partly due to the non-Borel summability of the theory because of  the existence of
classical soliton solutions. In this paper, we   offer a coherent formula for the
effective potential of the $PT$ symmetric and non Hermitian $\phi^{4}_{1+1}$
theory which is reliable for any coupling value. First, we study the effective
potential (at zero temperature) of the model in Eq.(\ref{fielng}) in an
effective quasi-particle theory which verify perturbation for both $g \ll1
\ \text{and}\ g \gg1$. For the intermediate region in the coupling space, the
quasi-particle theory ought to be non-perturbative and to resum the
perturbation series one has to resort to a resummation technique rather than
Borel technique because the theory is not Borel summable. Pade approximation
are suggested for the sake of getting reliable results from the input
information of perturbation series. However, the knowledge of only a few first
terms does not permit one to use these techniques \cite{Yukalov1}. This means
that to provide a reasonable accuracy, these techniques need to know tens of
first terms of the perturbation theory \cite{Yukalov2}. In fact, going to
higher orders in quantum field models is not an easy task as time-ordering of
many fields results in many different types of Feynman diagrams and thus one
needs (if it is possible to do the calculations) a long time to accomplish the
diagrams calculations. To overcome such difficulties, the self-similar method
was suggested as a non-perturbative tool for the resummation of divergent
series
\cite{Yukalov1,Yukalovb,Yukalov2,Yukalov3,Yukalov4,Yukalov5,Yukalov6,Yukalov8,Yukalov9}%
. Although this method can give good results even with few terms of
perturbation series, sometimes the method is not applicable at all. We will
argue, later in this work, it's applicability and suggestions for  modification to render it applicable for the effective potential for any coupling value.

The paper is organized as follows. In Section \ref{normal3}, we obtain the first order effective potential by normal ordering of the fields with a field as well as mass shift. Then, the first order calculations is supplemented by perturbative corrections up to $g^3$. In Section \ref{fraself}, we briefly review
 the key points of the fractal self-similar method and introduce more control functions to render it applicable for the effective potential for all the values of the coupling $g$. In Section \ref{numerical}, we present and discuss the results and Conclusion follows in Section\ref{conclusion}. 

\section{The perturbative effective potential \label{normal3}}

In low dimensional super-renormalizable theories, it is often enough to work
with normal ordering to render the quantum field theory finite. This is
because there are only few diagrams that are divergent and these are regulated
by normal ordering. The $\left(  \frac{-g}{4}\phi^{4}\right)  _{1+1}$ theory
is such an example that has only one divergent diagram in the self-energy
amplitude. In that case, one shall start with a Hamiltonian that is normal
ordered with respect to the vacuum of mass parameter $m$.
\begin{equation}
H=N_{m}\left(  \frac{1}{2}\left(  \left(  \nabla\phi\right)  ^{2}+\pi
^{2}+m^{2}\phi^{2}\right)  -\frac{g}{4}\phi^{4}\right)  . \label{hamilt2}%
\end{equation}
We can use the relation \cite{Coleman}
\begin{equation}
N_{m}\exp\left(  i\beta\phi\right)  =\exp\left(  -\frac{1}{2}\beta^{2}%
\Delta\right)  N_{M=\sqrt{t}\cdot m}\exp\left(  i\beta\phi\right)  \text{,}
\label{normal2}%
\end{equation}
to rewrite the Hamiltonian normal ordered\ with respect to a new mass
parameter $M=\sqrt{t}\cdot m$. In eq.(\ref{normal2}), expanding both sides and
equating the coefficients of the same power in $\beta$ yields the result%

\begin{align}
N_{m}\phi &  =N_{M}\phi,\nonumber\\
N_{m}\phi^{2}  &  =N_{M}^{2}\phi+\Delta,\nonumber\\
N_{m}\phi^{3}  &  =N_{M}\phi^{3}+3\Delta N_{M}\phi,\label{normall}\\
N_{m}\phi^{4}  &  =N_{M}\phi^{4}+6\Delta N_{M}\phi^{2}+3\Delta^{2},\nonumber
\end{align}

with
\begin{equation}
\Delta=-\frac{1}{4\pi}\ln t.
\end{equation}

Also, it is easy to obtain the result \cite{efbook}

\begin{equation}
N_{m}\left(  \frac{1}{2}\left(  \nabla\phi\right)  ^{2}+\frac{1}{2}\pi
^{2}\right)  =N_{M}\left(  \frac{1}{2}\left(  \nabla\phi\right)  ^{2}+\frac
{1}{2}\pi^{2}\right)  +\frac{1}{8\pi}\left(  M^{2}-m^{2}\right)  .
\label{norkin1}%
\end{equation}
The mass shift $m \rightarrow M$ should be accompanied by the canonical
transformation\cite{efbook}%

\begin{equation}
\left(  \phi,\pi\right)  \rightarrow\left(  \psi+B,\Pi\right)  .
\end{equation}
The field\ $\psi$ has mass $M=\sqrt{t}\cdot m$ , $B$ is a constant, the field
condensate and $\Pi$ is the conjugate momentum ($\overset{\cdot}{\psi}$).
Therefore, the Hamiltonian in Eq.(\ref{hamilt2}) can be written in the form;

\bigskip%
\begin{equation}
H=\bar{H}_{0}+\bar{H}_{I}+\bar{H}_{1}+E, \label{quasi}%
\end{equation}
where
\begin{align*}
\bar{H}_{0}  &  =N_{M}\left(  \frac{1}{2}\left(  \Pi^{2}+\left(
\triangledown\psi\right)  ^{2}\right)  \right)  +\frac{1}{2}N_{M}\left(
m^{2}-3g\left(  B^{2}+\Delta\right)  \right)  \psi^{2},\\
\bar{H}_{I}  &  =\frac{-g}{4}N_{M}\left(  \psi^{4}+4B\psi^{3}\right)  .
\end{align*}
$\bar{H_{1}}$ can be found as
\begin{equation}
\bar{H}_{1}=N_{M}\left(  m^{2}-g\left(  B^{2}+3\Delta\right)  \right)  B\psi,
\end{equation}
and the field independent terms can be regrouped as
\begin{equation}
E=\frac{1}{2}\left(  m^{2}-\frac{12g\Delta}{4}\right)  B^{2}-\frac{g}{4}%
B^{4}+\frac{1}{8\pi}\left(  M^{2}-m^{2}\right)  -\frac{3g\Delta^{2}}{4}%
+\frac{1}{2}m^{2}\Delta.
\end{equation}
Taking $b^{2}=4\pi B^{2}$ and the dimensionless parameters $t=\frac{M^{2}%
}{m^{2}}$, $G=\frac{g}{2\pi m^{2}}$ , the corresponding vacuum energy density
can be written as%
\begin{equation}
E(b,t,G)=\frac{m^{2}}{8\pi}\left(  b^{2}-\frac{G}{4}\left(  b^{4}-6b^{2}\ln
t+3\ln^{2}t\right)  +t-1-\ln t\right)  \label{effor}%
\end{equation}

The renormalization conditions are given by \cite{Peskin}%

\begin{equation}
\label{stab}\frac{\partial^{n}}{\partial b^{n}}E(b,t,G)=g_{n}\text{,}%
\end{equation}
where $g_{n}$ is the $\psi^{n}$ coupling. For instance,%
\begin{equation}
\text{\ \ }\frac{\partial^{2}E}{\partial B^{2}}=M^{2}\text{,} \label{ren}%
\end{equation}
where $g_{2}=M^{2}=m^{2}-3g\left(  B^{2}+\Delta\right)  $. Note that, the
renormalization condition $\frac{\partial E}{\partial B}=0$ enforces $\bar
{H}_{1}$ to be zero.

The quasi-particle Hamiltonian in Eq.(\ref{quasi}) exhibits  some interesting
properties. For instance, the renormalization conditions predicts an imaginary
condensate which turns the Hamiltonian to be non-Hermitian. To show how this
comes out, consider the equations%
\begin{align}
\frac{\partial E}{\partial b}  &  =\left(  2+\left(  -G\right)  \left(
b^{2}-3\ln t\right)  \allowbreak\right)  b=0,\\
\frac{\partial^{2}E}{\partial b^{2}}  &  =\left(  -G\right)  \left(
3b^{2}-3\ln t\right)  +\allowbreak2=2t.
\end{align}

For $b\neq0,$ they simplify to%

\begin{align}
2-G\left(  b^{2}-3\ln t\right)  \allowbreak &  =0,\nonumber\\
-G\left(  3b^{2}-3\ln t\right)  +\allowbreak2  &  =2t. \label{or1p1}%
\end{align}

Now, \bigskip Eq.(\ref{or1p1}) can be parameterized as%

\begin{align}
G  &  =-\frac{t+2}{3\ln t},\nonumber\\
b^{2}  &  =-\frac{t}{G}. \label{param}%
\end{align}
This parameterization shows that the parameter $t$ should be less than one and
$b$ is pure imaginary. This result clearly shows the non-Hermitian property of
the theory considering the form of $\bar{H}_{I}$ in Eq.(\ref{quasi}). 

The solutions of Eq.(\ref{or1p1}) are given by
\begin{align}
b^{2}(G)  &  =-3W\left(  \frac{1}{3G}e^{-\frac{2}{3G}}\right)  ,\label{lambb}%
\\
t(G)  &  =3GW\left(  \frac{1}{3G}e^{-\frac{2}{3G}}\right)  \allowbreak,
\label{lambt}%
\end{align}
where the Lambert's $\mathop{\rm W}(x)$ function is defined by $W(x)e^{W(x)}%
=x$. Also, $W$ has the series expansion \cite{complex}%
\begin{equation}
W\left(  z\right)  =%
{\displaystyle\sum_{n\geq1}}
\frac{\left(  -n\right)  ^{n-1}z^{n}}{n!},
\end{equation}
which is convergent if $\left\vert z\right\vert <\frac{1}{e}$. Thus, for
$G\rightarrow0^{+}$, Eq.(\ref{lambb}) takes the form
\begin{align}
b^{2}(G)  &  \simeq-3\left(  \frac{1}{G}e^{\frac{-2}{3G}}\right)
,\label{vanish}\\
b  &  =\pm i\sqrt{3}\left(  \frac{1}{\sqrt{G}}e^{\frac{-1}{3G}}\right)  .
\end{align}
This result predicts pure imaginary condensates which vanishes exponentially
as $G\rightarrow0^{+}$, which agrees with the prediction of Refs.
\cite{bendg1,bound}. Another interesting property that the quasi-particle
Hamiltonian posses is that it is totally equivalent to the original theory in
the sense that setting the parameter $t$ equal to one ($M=m$), the Hamiltonian $H$
(Eq.(\ref{quasi})) reproduces the original form in Eq.(\ref{hamilt2}). In fact, 
this is very important because, as we will see later in the work, the direct
calculation of higher orders, spoils out this equivalence. Also, since normal
ordering can not account for non-cactus Feynman diagrams, we resort to
renormalization group invariance to fix the parameters in the theory which
automatically turn the quasi-particle Hamiltonian equivalent to the original
Hamiltonian. Besides, it is interesting to note that Eq.(\ref{or1p1}) can be
obtained by adding a counter term as well as employing the invariance of the
bare couplings on the change of the mass scale $t$ \cite{ptsym1f}.

In spite of all of the above correct features, the normal ordered effective potential in Eq.(\ref{effor}), as we will discuss, is non-perturbative for intermediate values of the coupling \ $G$. In order to improve the
representation of the effective potential near the non-perturbative region, we
consider the modification of Eq.(\ref{effor}) resulting from the higher order
perturbative corrections to the vacuum energy followed by a modified fractal
self-similar method (the theory is not Borel summable).

\begin{figure}[th]
\begin{center}
\includegraphics{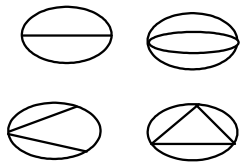}
\end{center}
\caption{The vacuum diagrams (up to $g^{3}$) of the effective quasi-particle
$(-\frac{g}{4}\phi^{4}-g B\phi^{3})$ theory.}%
\label{feyngk}%
\end{figure}

The normal ordered effective potential of $\frac{g}{4}\phi^{4}$ theory
(Eq.(\ref{effor})) agrees with GEP results \cite{gep4} which in turn accounts
not only for the leading order diagrams but also for all the non-cactus
diagrams \cite{wen-Fa,changcac}. Thus, to go to higher orders we include only
non-cactus diagrams (Fig.\ref{feyngk}).

In the equivalent quasi-field theory, the interaction term is $\left(
-\frac{g}{4}\phi^{4}-gB\phi^{3}\right)  $. Up to $g^{3}$, we have the Feynman
diagrams (non-cactus) shown in Fig.\ref{feyngk}. Accordingly, the perturbation
corrections to the effective potential are
\begin{align}
\label{epert}\frac{8\pi E(b,t,G)}{m^{2}}  &  =t-\ln t+b^{2}-1-G\left(
\frac{1}{4}b^{4}+ \frac{3}{4}\ln^{2}t-\frac{3}{2}b^{2}\ln t\right) \\
& +G^{2}\left( -\frac{3.155}{t}-3.515\frac{b^{2}}{t}\right)  -G^{3}\left(
\frac{4.057}{t^{2}}+9.918\frac{b^{2}}{t^{2}}\right) .\nonumber
\end{align}
In fact, this form of effective potential does not predict the values of $b$
and $t$ parameters that makes the quasi-particle theory equivalent to the
original theory. To keep the equivalence, we use the fact that the bare
parameters are independent of the scale $t$ \cite{collinbook}. Accordingly, we obtain the
result
\begin{align}
\frac{8\pi E(t,b,G)}{m^{2}}  &  =t-\ln t+b^{2}-1-G\left(  \frac{1}{4}%
b^{4}+\frac{3}{4}\ln^{2}t-\frac{3}{2}b^{2}\ln t\right) \nonumber\\
&  +G^{2}\left(  -3.155\left(  \frac{1}{t}-1\right)  -3.515b^{2}\left(
\frac{1}{t}-1\right)  \right) \label{Eperrg}\\
&  -G^{3}\left(  4.057\left(  \frac{1}{t^{2}}-1\right)  +9.918b^{2}\left(
\frac{1}{t^{2}}-1\right)  \right) \nonumber
\end{align}
We employ the conditions in Eq.(\ref{stab}) to predict the parameters $b$ and
$t$ which define the vacuum. This leads to the following conditions on the
parameters $b$ and $t$ (for $b\neq0$) :
\begin{align}
2-7.\,\allowbreak03G^{2}\left(  \frac{1}{t}-1\right)  -19.\,\allowbreak
836G^{3}\left(  \frac{1}{t^{2}}-1\right)  -G\left(  b^{2}-3\ln t\right)
\allowbreak &  =0\label{ORE1g3}\\
-19.\,\allowbreak836G^{3}\left(  \frac{1}{t^{2}}-1\right)  -7.\,\allowbreak
03G^{2}\left(  \frac{1}{t}-1\right)  -3G\left(  b^{2}-\ln t\right)
+\allowbreak2  &  =2t \label{ORE2g}%
\end{align}

In all of the $G^{3}$ order equations (Eq.(\ref{Eperrg}), Eq.(\ref{ORE1g3})
and Eq.(\ref{ORE2g})) setting $t=1$, we have $b=E=0$, which when substituted
back into Eq.(\ref{quasi}) we get the original Hamiltonian form in
Eq.(\ref{hamilt2}). Up to the best of our knowledge, such form of the
effective potential which conserve the equivalence between original theory and
the transformed one has never been obtained before (up to $g^{3}$ order).

As in Eq.(\ref{param}), \ Eq.(\ref{ORE1g3}) and Eq.(\ref{ORE2g}) can be
parametrized as

\bigskip$\allowbreak$
\begin{align}
t  &  =-3G\ln t+7.\,\allowbreak03G^{2}\left(  \frac{1}{t}-1\right)
-19.\,\allowbreak836{\left(  -G\right)  ^{3}\left(  \frac{1}{t^{2}}-1\right)
}-2,\nonumber\\
b^{2}  &  =-\frac{t}{G}. \label{orper}%
\end{align}
This parametrization is exactly the same as in Eq.(\ref{param}) except that
the $t$ parameter takes it's new form according to the radiative corrections.
One may conjecture, relying on this result, that the prediction of the vacuum
condensate as $b^2=\frac{-t}{G}$ is correct up to any order of the perturbation series and the
higher orders just redefine the parameters due to the radiative corrections.

As we will see in section \ref{numerical}, the $G^{3}$ order equations so
obtained has an interesting property. These equations are perturbative not
only for $G \ll1$, which is proved in Ref.\cite{bound} but also for $G \gg1$ (see Fig.\ref{eg},
Fig.\ref{bg} and Fig.\ref{tg}). To obtain an effective potential
which is reliable for all the coupling values and noting that the theory is
not Borel summable, we resort to the fractal self-similar method
\cite{Yukalov1}. In fact, we will modify it via the introduction of more
control functions to make it applicable for all the regions in the coupling space.

\section{The modified fractal Self-Similar method}\label{fraself}

 To get finite values of a divergent series, the so-called
resummation methods are suggested. The most often used among such techniques
are the Borel summation \cite{berzin,kleinert} and the construction of Pade
approximants \cite{baker}, including the two-point \cite{baker1} and
multivalued \cite{baker2,baker3} Pade approximants. These techniques have many
known limitations. For instance, The Borel method can not be applied in case of the existence of classical soliton solutions \cite{cacu} like the model we study in this work. Also, Borel and Pade techniques require to have a number of perturbative terms which often are hard to get. Rather than that, the self-similar method
can give good approximation with few terms at hand. One of the basic ideas in
self-similar approximation theory is the introduction of control functions
which govern the evolution of an approximation dynamical system to be close to
a fixed point. To introduce control functions into a given asymptotic series,
one has to employ a transformation that include trial parameters. This
transformation has to simulate the self-similarity property hidden in the
given perturbative sequence. For power series, it looks natural to employ the
power-law transformations \cite{Yukalov10}. Since power laws are typical of
fractals \cite{mandel,kroger} the power-law transformation can also be called
the fractal transformation \cite{Yukalov10}. Accordingly, it may be more
plausible to introduce control functions via a fractal transform because it
satisfies the scaling relation
\[
\frac{P(\lambda x,s)}{p(\lambda x)}=\lambda^{s} \frac{P (x,s)}{p(\lambda x)},
\]
which is a typical of fractals, where $P(x,s)$ and $p(x)$ are the polynomials
defined below.

For a function $f(x)$, the fractal transform  is given by
\begin{equation}
F(x,s)=x^{s}f\left(  x\right)  ,
\end{equation}
and it's inverse transform is:
\begin{equation}
f\left(  x\right)  =x^{-s}F(x,s)\text{.}%
\end{equation}

Here we follow the work in Ref.\cite{Yukalov1} to present the key points of
the self-similar method.

Consider the series given by%
\begin{equation}
p_{k}=\sum_{k=0}^{k}a_{n}x^{_{n}}.
\end{equation}

Applying the fractal transform we get%
\begin{equation}
P_{k}\left(  x,s\right)  =x^{s}P_{k}\left(  x\right)  =\sum_{n=0}^{k}%
a_{n}x^{s+n}.
\end{equation}

Define the initial approximation $P_{0}\left(  x,s\right)  =a_{0}x^{s}=f$.
\ Solving for $x$ \ we get%
\begin{equation}
x\left(  f,s\right)  =\left(  \frac{f}{a_{0}}\right)  ^{\frac{1}{s}}.
\end{equation}
Then
\begin{equation}
y_{k}\left(  f,s\right)  =P_{k}\left(  x\left(  f,s\right)  ,s\right)
=\sum_{n=0}^{k}a_{n}\left(  \frac{f}{a_{0}}\right)  ^{\frac{n}{s+1}}.
\end{equation}
Note that, self similarity means that%
\begin{equation}
y_{k+p}\left(  f,s\right)  =y_{k}\left(  y_{p}\left(  f,s\right)  ,s\right)  .
\end{equation}

Considering the cascade $y_k$ as a dynamical system with the time as $k$, then the  cascade velocity is given by%

\begin{align}
\label{diff}y_{k}\left(  f,s\right)  -y_{k-1}\left(  f,s\right)   &
=\sum_{n=0}^{k}a_{n}\left(  \frac{f}{a_{0}}\right)  ^{\frac{n}{s+1}}%
-\sum_{n=0}^{k-1}a_{n}\left(  \frac{f}{a_{0}}\right)  ^{\frac{n}{s+1}%
}\nonumber\\
&  =a_{k}\left(  \frac{f}{a_{0}}\right)  ^{\frac{n}{s+1}}.
\end{align}

After introducing the control functions, the regime of the self-similar
renormalization is to consider the the passage from one approximation to
another as a motion with respect to the approximation number $\ k=0,1,2$, ..
.. In fact, the trajectory $y_{k}\left(  f,s\right)  $ of this dynamical
system is bijective, that is, in one-to-one correspondence to the
approximation sequence $P_{k}\left(  x,s\right)  $. This dynamical system with
discrete time $k$ is called the approximation cascade. The attracting fixed
point of the cascade trajectory is, by construction, bijective to the limit of
the approximation sequence $P_{k}\left(  x,s\right)  $, that is, it
corresponds to the sought function.

One can deal with continuous time $t$ rather than the discrete time $k$ such
that the trajectory passes through the same points when $t=k$. In this case,
the flow velocity is governed by a differential equation rather the difference
equation in Eq.(\ref{diff}). In other words, the evolution equation for the
flow reads%
\begin{equation}
\frac{\partial}{\partial t}y\left(  t,f,s\right)  =v\left(  y\left(
t,f,s\right)  \right)  .
\end{equation}
Accordingly, the evolution integral is
\begin{equation}
\int_{P_{k}}^{P_{k+1}^{\ast}}\frac{df}{v_{k+1}\left(  f,s\right)  }%
=t_{k}^{\ast}.
\end{equation}
Thus, the self-similar approximation is given by \cite{Yukalov11}%
\begin{equation}
p_{k}^{\ast}=p_{k-1}\left(  x\right)  \left(  1-\frac{ka_{k}}{sa_{0}%
^{1+\frac{k}{s}}}x^{k}p_{k-1}^{\frac{k}{s}}\left(  x\right)  \right)
^{\frac{-s}{k}},
\end{equation}
where $t^{*}_{k} =1$ when no restrictions are imposed on the series
\cite{Yukalov11}.

The applicability of the method is governed by the stabilizers%

\begin{equation}
\mu_{k}\left(  f\right)  =\frac{\partial}{\partial f}y_{k}\left(  f,s\right)
,
\end{equation}
or their images
\begin{equation}
m_{k}\left(  x,s\right)  =\mu_{k}\left(  P_{0}\left(  x,s\right)  ,s\right)  .
\end{equation}
The stability condition is given by%
\begin{equation}
\left\vert m_{k}\left(  x,s\right)  \right\vert <1.
\end{equation}

\bigskip For the series given above we have
\begin{equation}
m_{k}\left(  x,s\right)  =%
{\displaystyle\sum\limits_{n=0}^{k}}
\frac{a_{n}}{a_{0}}\left(  1+\frac{n}{s}\right)  x^{n}, \label{stabx}%
\end{equation}

For $k=3$, \ the stabilizers are given by%
\begin{equation}
m_{k}\left(  x,s\right)  =\frac{xa_{1}+2x^{2}a_{2}+3x^{3}a_{3}}{sa_{0}%
}\allowbreak+\frac{a_{0}+xa_{1}+x^{2}a_{2}+x^{3}a_{3}}{a_{0}},
\end{equation}

The most stable aproximant is obtained if $m_{k}\left(  x,s\right)  =0$, or
\begin{equation}
s=-\frac{xa_{1}+2x^{2}a_{2}+3x^{3}a_{3}}{a_{0}+x a_{1}+x^{2}a_{2}+x^{3}a_{3}}.
\end{equation}
If $m_{k}\left(  x,s\right)  $ does not have a positive root, then
$m_{k}\left(  x,s\right)  $ is a monotonic decreasing function of $s$.
Therefor, the minimum is given by
\begin{equation}
\left\vert m_{k}\left(  x,s\right)  \right\vert _{s\rightarrow\infty
}=\left\vert \frac{a_{0}+xa_{1}+x^{2}a_{2}+x^{3}a_{3}}{a_{0}}\right\vert .
\end{equation}

 If all $\left\vert m_{k}\left(  x,s\right)  \right\vert
_{s\rightarrow\infty}$ are less than one, the bootstrap formula given by%

\begin{equation}
p_{k}^{\ast}=a_{0}\exp\left(  \frac{a_{1}}{a_{0}}x\exp\left(  \frac{a_{2}%
}{a_{1}}x\exp\left(  \frac{a_{3}}{a_{2}}x\exp\left(  \frac{a_{4}}{a_{3}}%
x\exp\left(  \frac{a_{5}}{a_{4}}x.......\exp\left(  \frac{a_{k}}{a_{k-1}%
}x\right)  \right)  \right)  \right)  \right)  \right)
\end{equation}
represents the resummed series \cite{Yukalovb}. However, it is not guaranteed to have all
$\left\vert m_{k}\left(  x,s\right)  \right\vert $ less than one for
$s\rightarrow\infty$ and for every point of the argument $x$. In fact, this is
the situation in case of applying the fractal self-similar method to the
effective potential in Eq.(\ref{Eperrg}) where there are some coupling values
for which the bootstrap formula does not exist. So, instead of applying
the the fractal self-similar method directly to the series in Eq.(\ref{Eperrg}%
) we introduce another control function via a bijective transformation which
has the property of transforming the original series to another one for which
the bootstrap formula is applicable. Then, we apply the fractal self-similar
method to the transformed series and at the end we apply the inverse of the
transformation to get the resummation formula of the original series.

To test the modification we introduced, consider the Lambert $W$ function defined by%

\begin{equation}
W\left(  x\right)  \exp\left(  W\left(  x\right)  \right)  =x.
\end{equation}

The series expansion of $W\left(  1+x\right)  $ is
\begin{align}
\allowbreak W\left(  1+x\right)   &  \approx W\left(  1\right)  +\frac
{W\left(  1\right)  }{1+W\left(  1\right)  }x+\left(  -\frac{1}{2}\left(
W\left(  1\right)  \right)  ^{2}\frac{2+W\left(  1\right)  }{\left(
1+W\left(  1\right)  \right)  ^{3}}\right)  \allowbreak x^{2}\\
&  +\left(  \frac{1}{6}\left(  W\left(  1\right)  \right)  ^{3}\frac
{9+8W\left(  1\right)  +2\left(  W\left(  1\right)  \right)  ^{2}}{\left(
1+W\left(  1\right)  \right)  ^{5}}\allowbreak\right)  \allowbreak
x^{3}+O\left(  x^{4}\right)  .
\end{align}

At $x=3$, $W\left(  1+x\right)  =1.\,\allowbreak202\,2$ and the perturbative
result (up $x^{3}$) is $1.\,\allowbreak918\,9$. The error percent is
$\left\vert \frac{1.\,\allowbreak202\,2-1.\,\allowbreak918\,9}{1.\,\allowbreak
202\,2}\right\vert \%=\allowbreak59.6\,16$ $\%$.

Let us apply the transformation $\Upsilon\left(  \allowbreak W\left(
1+x\right)  \right)  =W\left(  1+x\right)  +c$, where $c$ is used as a control
function. Apply the fractal self-similar method to $\Upsilon\left(
\allowbreak W\left(  1+x\right)  \right)  $ and find $c$ which make all the
$\left\vert m_{k}\left(  x,s\right)  \right\vert _{s\rightarrow\infty}$ less
than one and then apply $\Upsilon^{-1}$ to the obtained bootstrap formula we
get the result $W\left(  1+x\right)  \approx1.\,1798$ with the error percent
$\left\vert \frac{1.\,\allowbreak202\,2-1.\,1798}{1.\,\allowbreak
202\,2}\right\vert 100\%=\allowbreak2.\,\allowbreak069\,7\%$. This result
indicates the success of the modification we introduced to the fractal
self-self-similar method which can be summarized as: instead of using one control
function $s$ we introduce two control functions $s$ and $c$. With
$s\rightarrow\infty$, we adopt $c$ to obtain stable approximant. In fact, this
trick is necessary because it results in a unique formula for the approximant
for all the values of the coupling $G$. Since it is well known that the
effective potential is the generating functional for all 1PI amplitudes,
unique formula for the effective potential makes it easy to obtain the
different amplitudes via analytic differentiation. 

We applied the modified
fractal-self similar method to the perturbation series of the effective
potential in Eq.(\ref{Eperrg}). As we will see in the following section, the
resummed series fits the perturbative data for regions where the perturbation
series is reliable ($G\ll1$ and $G\gg1$).

\section{Numerical calculations and discussions\label{numerical}}

In this section, we present the numerical calculations concerning the
perturbative as well as non-perturbative calculations for the vacuum properties of the
$\left(  \frac{-g}{4}\phi^{4}\right)  _{1+1}$ non-Hermitian field theory. The
non-Hermiticity of the theory is clear in it's quantum field version without
the employment of boundary conditions in the complex $x$ space. This can be
extracted from Eq.(\ref{orper}), where we realize that the
condensation is always pure imaginary and thus the from in Eq.(\ref{quasi}) shows that the
theory is non-Hermitian as well as PT symmetric.

\begin{figure}[ptb]

\begin{center}
\includegraphics{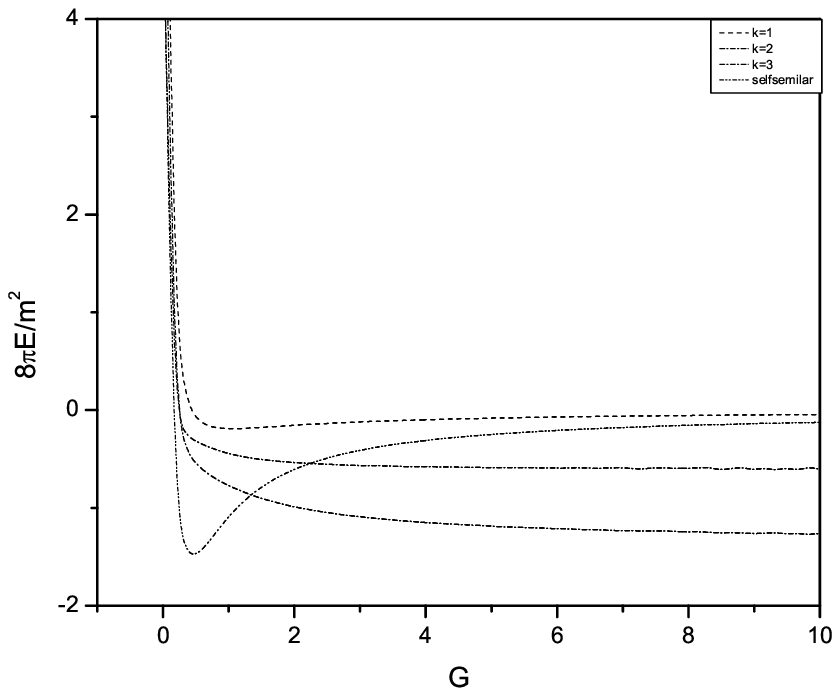}
\end{center}
\caption{The effective potential $8\pi\frac{E}{m^{2}}$ up to $G^{3}$ as a
function of the coupling .}%
\label{eg}%
\end{figure}

In Fig.\ref{eg}, the effective potential is plotted as a function of the
coupling $G$ for different values of order of perturbation $k$ as well as the
modified fractal-self similar resummation formula. Careful analysis of the
plot shows that the theory is really perturbative for both $G\ll1$ and $G\gg
1$. Although we can extract visually from the graph that the non-perturbative
calculations coincides for $k=1,2$ and $3$  for $G\ll1$, it is not clear that the
theory is perturbative for $G\gg1$ region. To clarify the prturbative
character of the theory for $G\gg1$, we rearrange Eq.(\ref{Eperrg}) as:

\bigskip%
\begin{equation}
\frac{8\pi E(t,b,G)}{m^{2}}=E_{b}+E_{0}\text{,}%
\end{equation}

where%

\begin{align}
E_{b} &  =b^{2}-G\left(  \frac{1}{4}b^{4}-\frac{3}{2}b^{2}\ln t\right)
  +G^{2}\left(  -3.515b^{2}\left(  \frac{1}{t}-1\right)  \right)\nonumber\\
&-G^{3}\left(  9.918b^{2}\left(  \frac{1}{t^{2}}-1\right)  \right)  \text{,}%
\end{align}

\begin{equation}
E_{0}=t-\ln t-1-G\left(  \frac{3}{4}\ln^{2}t\right)  +G^{2}\left(
-3.155\left(  \frac{1}{t}-1\right)  \right)  -4.057G^{3}\left(  \frac{1}%
{t^{2}}-1\right)  .
\end{equation}

\begin{figure}[ptb]
\begin{center}
\includegraphics{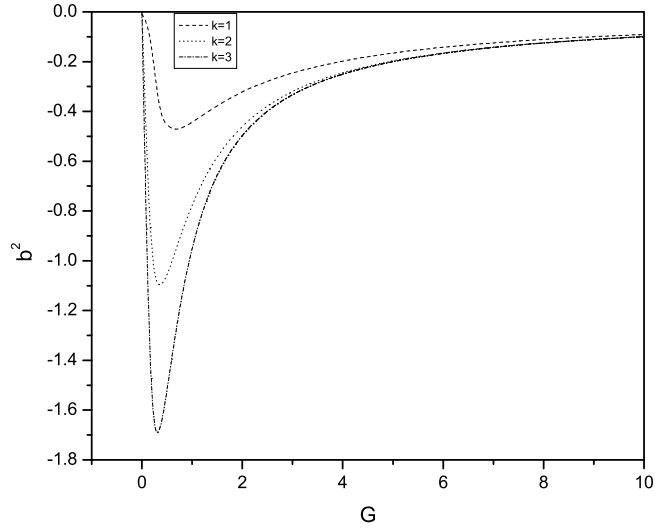}
\end{center}
\caption{The vacuum condensate squared versus the coupling $G$ for first,
second and third order in the perturbation series.}%
\label{bg}%
\end{figure}

In fact, the field dependent  term $E_{b}$ behaves well for both
$G\ll1$ and $G\gg1$. For $G\ll1$ it goes to zero as $G\rightarrow0$ \ while
$E_{0}$ goes to infinity. The appearance of infinite effective potential as
$G\rightarrow0$ can not be considered as Infra-Red divergence because it
appears only in the vacuum energy which can be rescaled as we can measure only
differences in energies. All the 1PI amplitudes are finite as $G\rightarrow0$
because we can get them by successive differentiation of the effective potential
with respect to the condensate $b$ \cite{Peskin} while $E_{0}$ has no
contribution. Thus, we
can safely consider the effective potential as $E_{b\text{ }}$ only. Also, when $G\gg1$ the theory is perturbative (see Fig.\ref{bg} and  Fig.\ref{tg}). 

\begin{figure}[th]
\begin{center}
\includegraphics[
height=3.2076in,
width=3.6737in
]{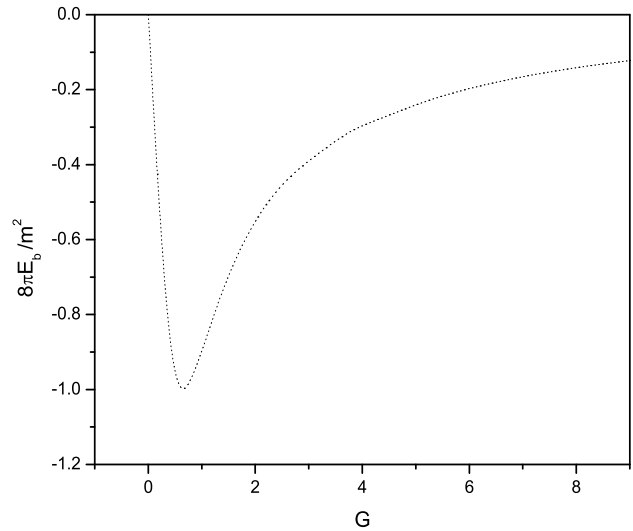}
\end{center}
\caption{The resummed vacuum energy $E_{b}$ as a function of the coupling $G$.}%
\label{egs}%
\end{figure}

\begin{figure}[th]
\begin{center}
\includegraphics[
height=3.2076in,
width=3.6737in
]{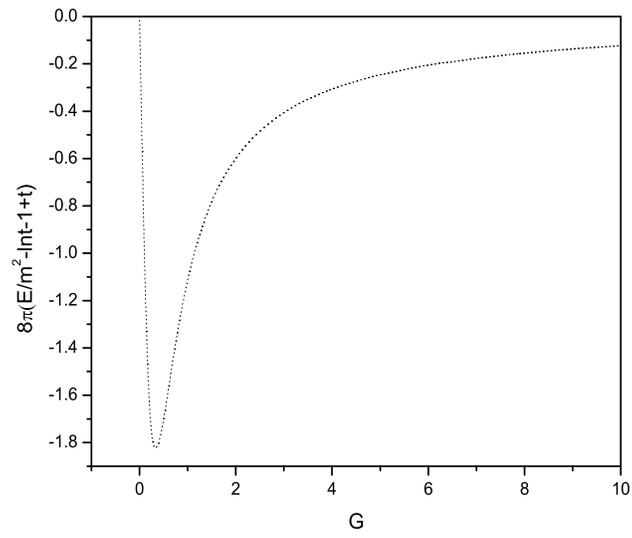}
\end{center}
\caption{The expectation  value of the potential term in the Hamiltonian as a function of the coupling $G$.}%
\label{egkt}%
\end{figure}

The fractal-self similar resummation for $E_b$ is plotted in Fig.\ref{egs} where we can realize that it goes to zero as $G\rightarrow 0$ and decreases as $G$ be very large. In fact, this is an indication that as we increase the coupling $G$ the number of bound states decreases as $G$ increases which is proved before in Ref.\cite{bound}. To assure this point, we plot the expectation value of the potential term in Fig.\ref{egkt} which shows that the depth of the potential decreases for large $G$ values and thus it is expected to have no bound states for $G\gg 1$.

\section{conclusion}

\label{conclusion} 

We investigated the effective potential of the $\left(  -\frac{g}{4}\phi
^{4}\right)  _{1+1}$ in view of the effective potential representation. The
normal ordered effective potential predicts a pure imaginary condensate but
tends to zero (exponential decrease) as $g\rightarrow0^{+}.$ The imaginary value of the
condensate turns the theory non-Hermitian without a priori postulated boundary
conditions which is necessary in the quantum  version of the theory. The
calculation of the effective potential is extended perturbatively up to
$g^{3}$ order which in turn predicts the same shape for the vacuum condensate
but with higher values. Up to $g^3$, we found that the theory is perturbative for both $g\ll1$ and $g\gg1$ regions.  Since the theory is ought to be non-perturbatieve for intermediate regions of the coupling space, we supplemented the perturbation result by the fractal-self similar method in a way that results in a coherent formula to make the calculation of all the 1PI amplitudes accessible analytically. 

We believe that these new results concerning the effective potential of the
$PT$ symmetric $\phi_{1+1}^{4}$ theory is a step forward toward the
understanding of these novel kind of field theories.

\begin{figure}[ptb]
\begin{center}
\includegraphics{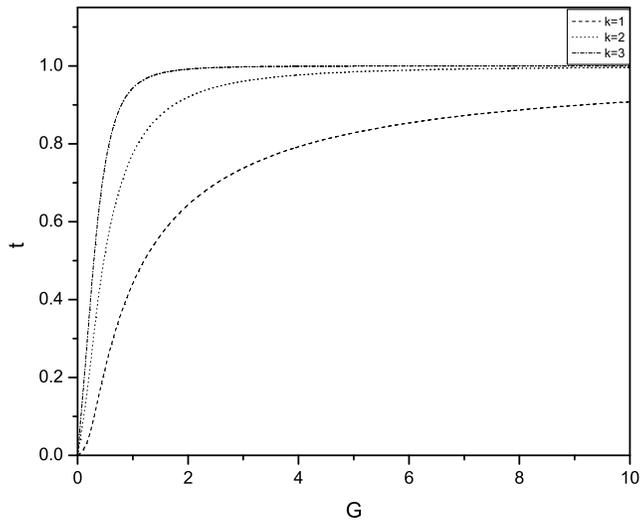}
\end{center}
\caption{The reciprocal of the 1PI two-point function versus the coupling $G$ for first,
second and third order in the perturbation series.}%
\label{tg}%
\end{figure}

A very interesting note should be mentioned, the simple model we used shows many similar features with \textbf{QCD}. For instance, In Ref.\cite{bound}, it was shown that the theory has two body bound state similar to $q\bar{q}$ and three body bound state like baryons. Also, our calculations shows that the theory is perturbative for large scales (Fig.\ref{tg}) which may be a clue that the theory may have an Ultra-Violet fixed point like \textbf{QCD}. Finally, the theory may have a symmetry restoration for large $G$ values. All of these expected features will be our task of the next work.

\section*{Acknowledgment}

\label{ack} The author would like to thank Dr. S.A. Elwakil for his support
and kind help. Also, deep thanks to Dr. C.R. Ji, from North Carolina State
University, for his direction to my attention to the critical phenomena in QFT
while he was \ supervising my Ph.D.

\newpage

\newpage

\end{document}